\documentstyle[twoside,fleqn,espcrc2,epsfig]{article}

\newcommand{\eq}{\begin{equation}}
\newcommand{\en}{\end{equation}}
\newcommand{\bea}{\begin{eqnarray}}
\newcommand{\eea}{\end{eqnarray}}

\newcommand{\bdm}{\begin{displaymath}}
\newcommand{\edm}{\end{displaymath}}

\newcommand{\qq}{{q\bar{q}}}

\newcommand{\ba}{\begin{array}}
\newcommand{\ea}{\end{array}}

\newcommand{\rn}{\scriptsize{{\frac{r}2}}}
\newcommand{\mrn}{\scriptsize{{-\frac{r}2}}}
\newcommand{\Rn}{\scriptsize{{\frac{R}2}}}
\newcommand{\mRn}{\scriptsize{{-\frac{R}2}}}
\newcommand{\ZZ}{\hbox{{\rm Z{\hbox to 3pt{\hss\rm Z}}}}}

\newcommand{\Br}{\langle}
\newcommand{\kt}{\rangle}

\newcommand{\uq}{\frac14}

\newcommand{\AmS}{{\protect\the\textfont2
  A\kern-.1667em\lower.5ex\hbox{M}\kern-.125emS}}

\title{Nature of the Vacuum inside the Color Flux Tube}

\author{F. Gliozzi\address{Dipartimento di Fisica Teorica, Universit\`a 
di Torino, \\ 
        via P. Giuria 1, 10125 Torino, Italy}
        \thanks{Speaker at the conference}
        and 
        S. Vinti$^a$}
       
\begin{document}

\begin{abstract}
The interior of the color flux tube joining a quark pair  
can be probed by evaluating the correlator of  pair of Polyakov loops
in a vacuum modified by another  Polyakov pair,
in order to check the dual superconductivity conjecture, which predicts 
a deconfined, hot core. We also point out  that at the critical point 
of any $3D$ gauge theories with a continuous deconfining transition 
the Svetitsky-Yaffe conjecture provides us with an analytic expression 
of the Polyakov correlator as a function of  the location of the probe inside 
the flux tube. Both these predictions are compared with numerical results in
$3D$ $Z_2$ gauge model, finding complete agreement.
\end{abstract}

\maketitle

\section{INTRODUCTION}

The underlying assumptions of the dual superconductivity\cite{tmp} of
gauge theories, and its appropriatenss for describing quark confinement, 
are not rigorously founded, and it is necessary  to perform precise numerical 
or analytic tests of this conjecture whenever possible.

The internal structure of the color flux tube joining a quark pair provides an
important test of these ideas, because it should show, as the dual of 
an Abrikosov vortex, a very peculiar property:  it is expected to have 
a  core of normal, hot vacuum as contrasted with the surrounding 
medium, which is in the dual superconducting phase. The location of the core 
would be given by the vanishing of the disorder parameter $\Br\Phi_M(x)\kt=0$, 
where $\Phi_M$ is some effective magnetic Higgs field. 

In a pure gauge theory, the formulation of this property from the
first principles poses some problems, because no local, gauge invariant, 
disorder field $\Phi_M(x)$ is known. As a consequence, one cannot define in a 
meaningful, precise way the notion of core of the dual vortex. 

A possible way out is suggested by the fact that in a medium in which 
$\Br\Phi_M\kt=0$ the quarks should be deconfined, then it is expected that 
the interquark potential inside the flux tube gets modified. As a consequence,
 one may try to define a gauge-invariant notion of normal core of the flux 
tube as the region where the interquark interaction mimics a deconfined 
behavior.
Of course one cannot speak of a true deconfinement, as it would require 
pulling infinitely apart the quarks, while the alleged core has a finite size.

A simple, practical way to study in a lattice gauge theory the influence
 of the flux tube on the quark interaction is based on the study of the 
system of four coplanar Polyakov loops $P_1,P_2,P_3$ and $P_4$ 
following two steps
\begin{description}
\item{~~} Modify the ordinary vacuum  by
inserting in the action the pair  $P_3,P_4^{\dagger}$
acting as sources at a fixed distance $R$.
\item{~~} Evaluate in this modified vacuum the correlator the other 
pair  $P_1,P_2^{\dagger}$ of Polyakov loops which are used  as probes.
\end{description}
The correlators in the two vacua are related by
\eq
\Br P_1 P_2^{\dagger}\kt_\qq=\frac{\Br P_1 P_2^{\dagger}P_3 P_4^{\dagger}\kt}
{\Br P_3 P_4^{\dagger}\kt}~~.
\en
In this  note we study some general properties of these correlators for 
$T\leq T_c$~. In particular, we point out that at $T=T_c$ the functional form 
of these correlators is universal and in some $3D$ gauge theories 
can be written explicitly, even in finite volumes.

\section{FOUR POLYAKOV LOOPS}
 
Consider the system of four parallel, coplanar Polyakov loops, symmetrically 
disposed with respect the origin  of a cubic lattice with periodic boundary 
conditions in the direction of the imaginary time (which coincides with the 
common direction of the loops). We study their correlator
\eq
\Br P_1 P_2^{\dagger}P_3 P_4^{\dagger}\kt=
\Br P(\mrn)P^{\dagger}(\rn)P(\mRn)P^{\dagger}(\Rn)\kt
\label{four}
\en
as a function of $r\le R$. 
For large $R$ and $r\sim R$ it obeys the asymptotic factorization 
condition
\eq
\Br P_1 P_2^{\dagger} P_3 P_4^{\dagger}\kt\sim
\Br P_1 P_3^{\dagger}\kt\Br P_2 P_4^{\dagger}\kt~.
\label{fact}
\en
When $T < T_c$, assuming the usual area law 
$\Br P_1P_2^{\dagger}\kt\propto\exp(-\sigma r/T)$,
where $\sigma$ is the string tension, yields
\eq
\Br P_1 P_2^{\dagger}\kt_\qq\sim\exp(\sigma r/T)
\sim1/\Br P_1 P_2^{\dagger}\kt~~,
\en
which gives an apparent repulsion between the two probes due to the 
attraction of the two sources.

The other limit $r\ll R$ is more interesting, because the kinematics does not
force any factorization and  different confinement models suggest
different behaviors. In particular in the naive string picture one is tempted
to assume the factorization (\ref{fact}) even in this limit, because within this
assumption the total area of the surfaces connecting the Polyakov loops is
minimal. On the contrary, in the dual superconductivity it is expected that 
the test particles probe the short distance properties of the 
hot core of the flux tube, thus the correlator in the modified vacuum would 
approach to a constant ($\sim \Br P\kt^2_{T>T_c}$) 
from above and
\eq
\Br P_1 P_2^{\dagger}\kt_\qq>\Br P_1 P_2^{\dagger}\kt~~~(r\ll R\,, T<T_c)~.
\en
In the range $T\ge T_c$ the interior of the flux tube is in the same phase of
the surrounding region and the mutual interaction between the two near 
probes should not depend on the presence of very far sources, then 
\eq
\Br P_1 P_2^{\dagger}\kt_\qq\sim\Br P_1 P_2^{\dagger}\kt~~~
(r\ll R\,, T\ge T_c)~.
\label{faq}
\en

\subsection{ Critical Behavior}

 According to the widely tested Svetitsky-Yaffe conjecture,  any gauge theory 
in $d+1$ dimensions with a continuous deconfining transition belongs to the 
same universality class of a $d$-dimensional $C(G)$-symmetric spin model, where 
$C(G)$ is the center of the gauge group. It follows that at the critical 
point all the critical indices describing the two transitions and all the 
adimensional ratios of correlation functions of corresponding observables 
in the two theories should coincide. 
In particular, since the order parameter the gauge theory is obviously
mapped in the corresponding one of the spin model, the correlation functions  
among  Polyakov loops should be proportional to the corresponding correlators 
of spin operators:
\eq
\Br P_1\dots P_{2n}\kt_{T=T_c}\propto \Br s_1\dots s_{2n}\kt~~.
\en
Conformal field theory has been very successful in determining 
the exact form of these universal functions for $d=2$ even in a finite box, 
which is a precious information for a correct comparison with numerical 
simulations.

In particular, using the known results of the $2D$ critical Ising model 
in a rectangle $L_1\times L_2$ with periodic boundary conditions \cite{fsz} 
we can write explicitly the correlator of  any (even) number $2n$ of 
Polyakov loops of any $2+1$ gauge theory with  $C(G)=\ZZ_2$. Let $x_j,y_j$
be the spatial coordinates of $P_j$ and define the complex variables 
$z_j=\frac{x_j}{L_1}+i\frac{y_j}{L_2}$ and $\tau=iL_2/L_1$. Then
\eq
\Br P_1\dots P_{2n}\kt^2=c_n\sum_{\nu=1}^{4}
\sum_{\varepsilon_i=\pm1}^{~}{\,}'
A_\nu(\varepsilon \cdot z)\prod_{i<j}B_{ij}
\label{crt}
\en
with $\varepsilon\cdot z=\sum_i \varepsilon_iz_i$ and the primed sum 
is constrained by $\sum_i\varepsilon_i=0$~; $c_n$ is an overall constant 
that can be expressed by factorization in terms of $c_1$.
The universal functions 
$A_\nu$ and $B_{ij}$ can be written in terms of the four Jacobi theta
functions $\vartheta_\nu(z,\tau)$ as follows 
\eq
B_{ij}=\left\vert\frac{\vartheta_1(z_i-z_j,\tau)}{\vartheta_1'(0,\tau)}
\right\vert^{\varepsilon_i\varepsilon_j/2},
\en
\eq
A_\nu(z)=\left\vert\frac{\vartheta_\nu(z,\tau)}{\vartheta_\nu(0,\tau)}
\right\vert^2,
\en

In the infinite box limit $L_1,L_2\to\infty$, using the Taylor expansion
\eq
\vartheta_\nu(z,\tau)=a_\nu(1-\delta_{1,\nu})+b_\nu\,z+O(z^2)~,
\en
the correlator (\ref{four}) becomes
\eq
\Br P_1P_2P_3P_4\kt=\frac{4c_1^2}{(Rr)^\uq}
\sqrt{\frac{R+r}{R-r}}~~,
\en
which satisfies both factorizations (\ref{fact},\ref{faq}).

\section {CLUSTER ALGORITHM}

In order to test the above formulae at criticality it is convenient to perform
the numerical simulations in the simplest model belonging to the 
above-mentioned universality class, which is the the $3D$ $\ZZ_2$ gauge model.  

Using the duality transformation it is possible to build 
up a one-to-one mapping of physical observables of the gauge system 
 into the corresponding spin quantities.
A great advantage of this method is that it can be used a  non local cluster 
updating algorithm \cite{sw}, which has been proven very successful in 
fighting critical slowing down. 

In this framework it is easily  shown that the vacuum expectation value of 
any set $\{C_1\dots C_n\}$ of  Polyakov or Wilson loops of arbitrary 
shapes  is simply  
encoded in the topology of Fortuin-Kasteleyn (FK) clusters: to each Montecarlo
configuration we assign a weight 1 whenever there is no FK cluster 
topologically linked to any $C_i\in\{C_1\dots C_n\}$, otherwise we assign a
weight 0. Let $N_0$ and $N_1$ be the number 
configurations of weight 0 and 1 respectively, then we have simply
\eq
\Br C_1\dots C_n\kt=\frac{N_1}{N_0+N_1}~~.
\en
This method provides us with a handy, very powerful tool to estimate 
the correlator of any set of Wilson or Polyakov
loops even at criticality.

\section{RESULTS}

In order to test the critical behavior of the multiloop correlator one has to
know with high precision the location of the critical temperature as a function 
of the coupling $\beta$. We took advantage of ref.\cite{ch}, where these
critical values have been obtained with an extremely high statistical 
accuracy. We report  in Fig.1 some results at $\beta=0.746035$ corresponding 
to $1/aT_c=N_{tc}=6 $
and to a string tension $\sigma a^2=0.0189(2)$. 
The open circles are the data for the correlator 
$\Br P(\mrn)P(\rn)\kt$ in a $N_t\times N_x\times N_y$ lattice with 
$N_t=3N_{tc},N_x=N_y=64$.
 They are well fitted by the one-parameter formula
$c\exp(-\sigma N_t r)/\eta(i\frac{N_t}{2r})$, where the Dedekind $\eta$
function takes into account the quantum contribution of the flux tube 
vibrations \cite{pv}.
The square symbols correspond to the correlator $\Br P(\mrn)P(\rn)\kt_\qq$ 
at the same temperature, in presence of  a pair of sources  at a distance 
$R=24a$. The data in the 
central region are well fitted by a two-parameter formula 
$c_\qq\exp(-\sigma N_t r)/\eta(i\frac{N_t}{2r})+b\to
c'_\qq\frac{e^{-mr}}{\sqrt{r}}+b$ which simulates a high temperature
behavior with a screening mass $m=\sigma N_t$ and an order parameter 
$\Br P\kt=\sqrt{b}$.  
The black circles correspond to
       $\Br P(\mrn)P(\rn)P(\mRn)P(\Rn)\kt$ evaluated at $T=T_c$ with $R=16a$. 
They fit nicely eq.(\ref{crt}) (continuous line). Note that such a curve 
does not contain any free parameters, being $c_2=\sqrt{2}c_1^2$ with 
$c_1 N_x^{\uq}=0.199(4)~$ as estimated by measuring $\Br P_1P_2\kt$
on lattices of different sizes at $T=T_c$ and $N_{tc}=6$.
%%%%%%%%%%%%%%%%%%%%%%%%%%%%%%%fig1%%%%%%%%%%%%%%%%%%%%%%%%%%
\vskip0.3cm
\vskip -2.3 cm
\hskip-2.cm\epsfig{file=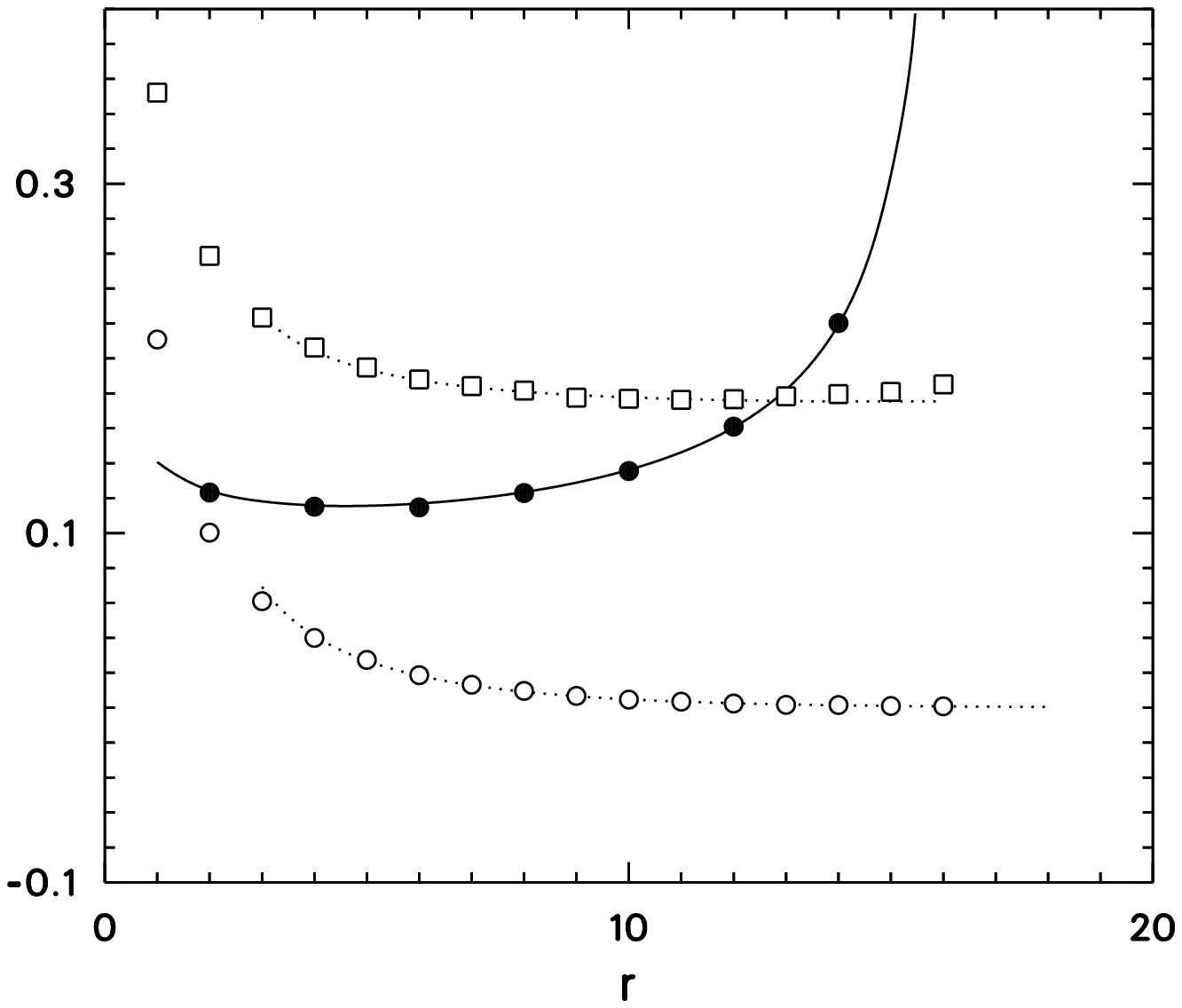,height=10.5cm}
%\vskip .02cm
\vskip -1.9cm
Figure. 1. { Correlator of two Polyakov loops inside and outside 
the flux tube}
\vskip0.2cm
%%%%%%%%%%%%%%%%%%%%%%%%%%%%%%%%%%%%%%%%%%%%%%%%%%%%%%%%%%%%

\end{document}